\documentclass[]{aa}

\usepackage{natbib}
\bibpunct{(}{)}{;}{a}{}{,} 

\usepackage{graphics}   
\usepackage{graphicx}   
\usepackage{epstopdf}
\usepackage{longtable}   
\usepackage{url}      
\usepackage{bm}        

\usepackage[varg]{txfonts}
\usepackage{pdflscape}
\usepackage{supertabular}
\usepackage{hyperref}
\usepackage{xcolor}
\usepackage[normalem]{ulem}

\definecolor{green}{rgb}{0.3,0.7,0.}

\newcommand\gva{{\sc Genec}}

\hypersetup{
    bookmarks=true,         
    unicode=false,          
    colorlinks=true,       
    linkcolor=blue,          
    citecolor=blue,        
    filecolor=blue,      
    urlcolor=blue           
}

\usepackage{color}

\newcommand{\msun}{\ensuremath{\mathrm{M}_{\odot}}\xspace}

\begin{document}

\title{The evolution and impact of $\sim$ 3000 M$_{\odot}$ stars in the early Universe}

\author{D. Nandal\inst{1}, E. Farrell\inst{1}, G. Buldgen\inst{1,2}, G. Meynet\inst{1}\and S. Ekström\inst{1}}

\institute{D\'epartement d'Astronomie, Universit\'e de Gen\`eve, Chemin Pegasi 51, CH-1290 Versoix, Switzerland \and STAR Institute, Université de Liège, Liège, Belgium}

\date{January, 2023}

\abstract{
We present evolutionary models of massive, accreting population III stars with constant and variable accretion rates until the end of silicon burning, with final masses of $\sim 1000 - 3000\ \msun$.
In all our models, after the core-hydrogen-burning phase, the star expands towards the red side of the Hertzsprung-Russell diagram is where it spends the rest of its evolution. 
During core helium burning, the models exhibit an outer convective envelope as well as many large intermediate convective zones. 
These intermediate zones allow for strong internal mixing to occur which enriches the surface in helium. The effect of increasing metallicity at a constant accretion rate of 10$^{-3}$ M$_\odot$yr$^{-1}$ shows an increase in the lifetime, final mass and distribution of helium in the envelope. Our fiducial model with mass of 3000 \msun has a final surface helium abundance of 0.74 and 9\% of its total mass or 50\% of the core mass, has a value of $\Gamma_1 < 4/3$ at the end of core silicon burning. If the collapse of the core is accompanied by the ejection of the envelope above the carbon-oxygen core, this could have a significant impact on the chemical evolution of the surroundings and subsequent stellar generations.
The model has a final log(N/O) $\approx 0.45$, above the lower limit in the recently detected high-redshift galaxy GN-z11.
We discuss the impact of a single 3000 \msun star on chemical, mechanical and radiative feedback, and present directions for future work. 

}


\titlerunning{Massive stars as Nitrogen sources}
\authorrunning{Devesh Nandal et al.}

\maketitle

\section{Introduction}

Population III (Pop III) stars, having masses in excess of 1000 $\msun$, are theorised to have had a multifaceted influence on the early Universe. They may contribute to the production of heavy elements via supernova explosions \citep{Nomoto2006}.
Accreting zero- and low-metallicity $1000 \msun$ stars could undergo core collapse, forming intermediate mass black holes, which would be seeds for the formation of supermassive black holes \citep{Ohkubo2009}.
Additionally, such objects have been invoked to explain recent observations of strong nitrogen enrichment in the high-redshift galaxy GN-z11 \citep{Bunker2023,Cameron2023, Charbonnel2023}.
In this work we compute evolutionary models of accreting Pop III stars to investigate their impact on chemical, mechanical, and radiative feedback in the early Universe.

Simulations of the collapse of primordial molecular clouds suggest that Pop III stars could grow to around 1000 - 10,000 M$_\odot$ \citep{Larson2000,Regan2009,Regan_2022}. If the primordial cloud consists of atomically cooled hydrogen instead, the final mass of the resulting objects would exceed 100,000 M$_\odot$ and they would collapse via general relativistic instability \citep{Hosokawa_2013,Lionel2018}. With such a wide range of final masses, Pop III stars are also expected to have a wide variety of impacts \citep[e.g.][]{Hoyle1964,Feitzinger1980,Cassinelli1981,Denissenkov2014}.
For instance, stellar objects with masses of around 10$^5$-10$^6$ M$_\odot$ could be strong sources of helium in the early Universe \citep{Hoyle1964}. 
Such supermassive stars (SMSs) may have an impact on the $^{26}$Al enrichment of the Milky Way \citep{Hille1987} and are thought to be the progenitors of the supermassive black holes currently observed at high redshifts, z>6 \citep{Haem2020}. 

\citet{Ohkubo2009} found that accreting zero- and low-metallicity $1000 \msun$ stars could undergo core collapse and form intermediate mass black holes, which would be seeds for the formation of supermassive black holes.
\citet{Volpato2023} investigated the evolution of Pop III stars up to 1000 M$_\odot$ and find that they could produce black holes of 40 - 1000 \msun.
\citet{Ledoux1982} studied the stability of a 3000 M$_\odot$ star on the zero-age main sequence (ZAMS) using static models and found that such an object would be vibrationally unstable as a result of the onset of nuclear burning.

The growth of such massive objects is expected to occur via the cold disc accretion scenario in primordial environments with accretion rates $\approx 10^{-3}M_{\odot}yr^{-1}$, sometimes exceeding $10^{-2}M_{\odot}yr^{-1}$, as shown in works by \citet{Yoshida2006}, \citet{Ohkubo2009}, \citet{Hirano2014}, \citet{Hosokawa2016}, and \citet{Chiaki2022}. A further offloading of matter onto the accretion disc may destabilise it and lead to fragmentation, thereby limiting the final mass of accreting Pop III stars \citep{Susa2019,Klessen2023}. Work by \citet{Chon2020} suggests that, despite fragmentation, the protostars may quickly merge to produce a single Pop III star with a mass $\approx 10^{5} \msun$. 

Directly observing such objects is currently beyond the reach of our instruments. Indirect measurements, in particular the detection of unusually strong N III] and N IV] UV emission lines in the spectrum of GN-z11 by the \textit{James Webb} Space Telescope (JWST), have elicited various proposed explanations that centre around the N/O abundance ratio in the nearby interstellar medium \citep{Bunker2023,Cameron2023}. One such scenario calls for a collision of massive stars with a SMS to produce a star with a mass of $10^{4} \msun$ \citet{Charbonnel2023}. Another scenario points towards metal-enriched SMSs  ($0.1\times Z_{\odot}$) producing the necessary abundances by means of stellar winds and explosions \citep{Nagele2023}. 

In this work we compute evolutionary models of Pop III stars with variable and constant accretion rates until the end of core silicon burning.
We show that, due to extended outer and intermediate convective zones appearing during the core-He-burning phase, the envelopes are significantly enriched in helium and nitrogen during the later stages. If this mass is lost, it may significantly enrich the interstellar medium in helium and nitrogen elements. 
The paper is organised as follows: Our modelling approach is described in Sect.~\ref{sec:method}, along with a discussion of the impact of the assumption of a constant accretion rate. 
Section~\ref{Sec:Evolution} describes the evolution of our Pop III stellar models with a constant accretion rate. Section~\ref{Sec:Impact} discuss the chemical, mechanical, and radiative impact of such objects, compares our results with previous work, and attempts to explain the chemical signature of GN-z11 using our model. The effect of changing the metallicity of such objects is presented is Sect.~\ref{Sec:Metal}. Finally, the main conclusions are presented in Sect.~\ref{Sec:Conc}.

\section{Description of our models of accreting stars} \label{sec:method}

\subsection{Physical ingredients}

The model analysed in this study was computed using the Geneva Stellar Evolution code \citep{Eggenberger2008}. We used similar physical ingredients as \citet{Ekstrom2012} but with a primordial initial composition, as in \citet{Murphy2021}. To model accretion, we assumed the infall of matter occurs via a geometrically thin cold disc and the specific entropy of the accreted matter is the same as that of the stellar surface \citep{haemmerle2014,Lionel2016}. 
Implicitly, this assumes that any entropy excess in the infalling matter is radiated away before it falls on the stellar surface \citep{Palla1992, Hosokawa_2013}. Hydrodynamical models of Pop III star formation have suggested accretion rates ranging from $10^{-6}$ to $10$ M$_\odot$/yr \citep{OShea2007, Yoshida2007, Shingo2014, Hosokawa2016}. In this work we chose either a constant accretion rate of $10^{-3}$ M$_\odot$/yr or variable accretion rates motivated by the hydrodynamical simulations of \citet{Yoshida2007}.

\subsection{Assumption of a constant accretion rate}\label{Sec:CVar}

Studies that model accreting stars with masses $\gtrsim 10^{3} M_{\odot}$ often assume a constant accretion rate that represents the net balance of the infall of matter onto the star and various processes that induce mass loss \citep{Hosokawa_2013, Woods2017, Lionel2018}. 
However, hydrodynamic simulations concerning primordial star formation point towards the accretion rate varying as a function of time rather than remaining constant, due to dynamical interactions and feedback mechanisms that govern the star growth \citep{Chen_2014, Sakurai2015, Wise_2019, Regan_2022}.

\begin{figure}
        \centering
          \includegraphics[width=9.0cm]{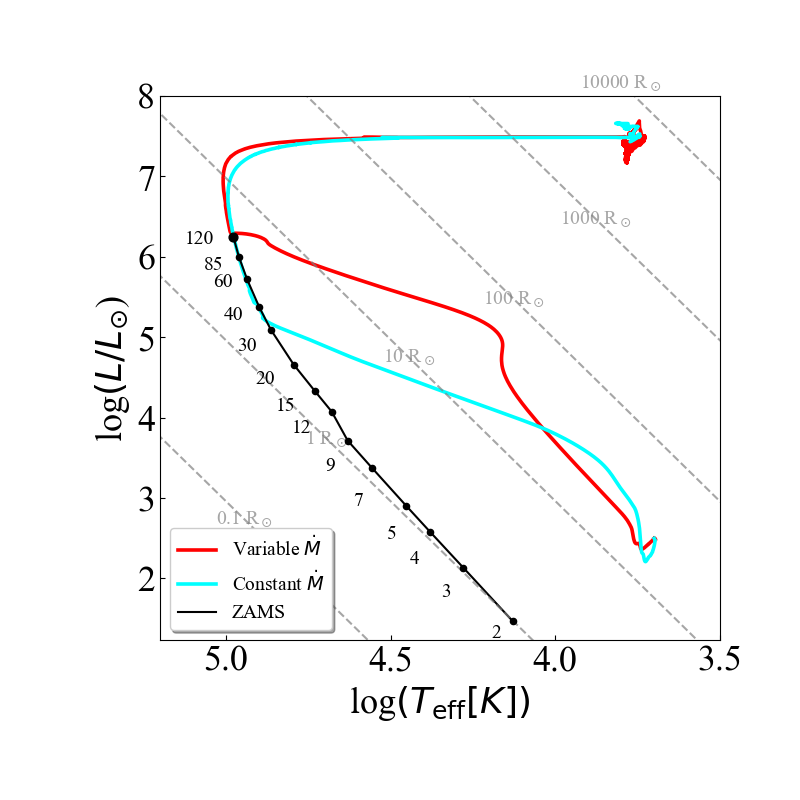}     
        \caption{HR diagrams of two models with identical masses of 917 M$_\odot$,  accreted until the end of core silicon burning. The red line depicts the model computed using the variable accretion rate from \citet{Yoshida2007}, and the cyan line represents a model with a constant accretion rate of $2.615x10^{-4} M_{\odot}/yr$. The black line is the ZAMS line, and the dots indicate the masses in M$_\odot$. The dotted grey lines are the iso-radius line, with values depicted in solar radii.}
                \label{Fig:HRD_convar}
\end{figure}

To test the impact of the assumption of a constant accretion rate on the properties of the final structure, we computed a model with a variable accretion rate from \citet{Yoshida2007} that is based on hydrodynamical simulations. For comparison, we also computed a model with a constant accretion rate ($2.62\times10^{-4} M_{\odot}/yr$), chosen such that it reaches the same final mass (917 M$_{\odot}$) as the model computed with the variable accretion rate from \citet{Yoshida2007}.
The variable accretion rate model has a higher accretion rate than the constant accretion rate model during the pre-main sequence.
As a result, the constant accretion rate model reaches the ZAMS with a mass of 34 M$_\odot$, compared to 120 M$_{\odot}$ for the variable accretion rate model (Fig.~\ref{Fig:HRD_convar}).
The constant accretion rate model has a main-sequence lifetime that is 11 \% longer, which may have an impact on ionising feedback.
The two models have a very similar evolution from core hydrogen burning until the end of core silicon burning. The radius and abundance profiles for the two models are also extremely similar, indicating that the chemical and radiative feedback of the two models are indistinguishable. 
Therefore, for the same final mass, the choice of accretion history does not significantly impact the final stellar structure.
As a result, we make the simple assumption of a constant accretion rate when discussing the detailed evolution in the following section.

\section{Evolution of an extremely massive accreting Pop III  star}\label{Sec:Evolution}

\begin{figure*}
        \centering
                \includegraphics[width=9cm]{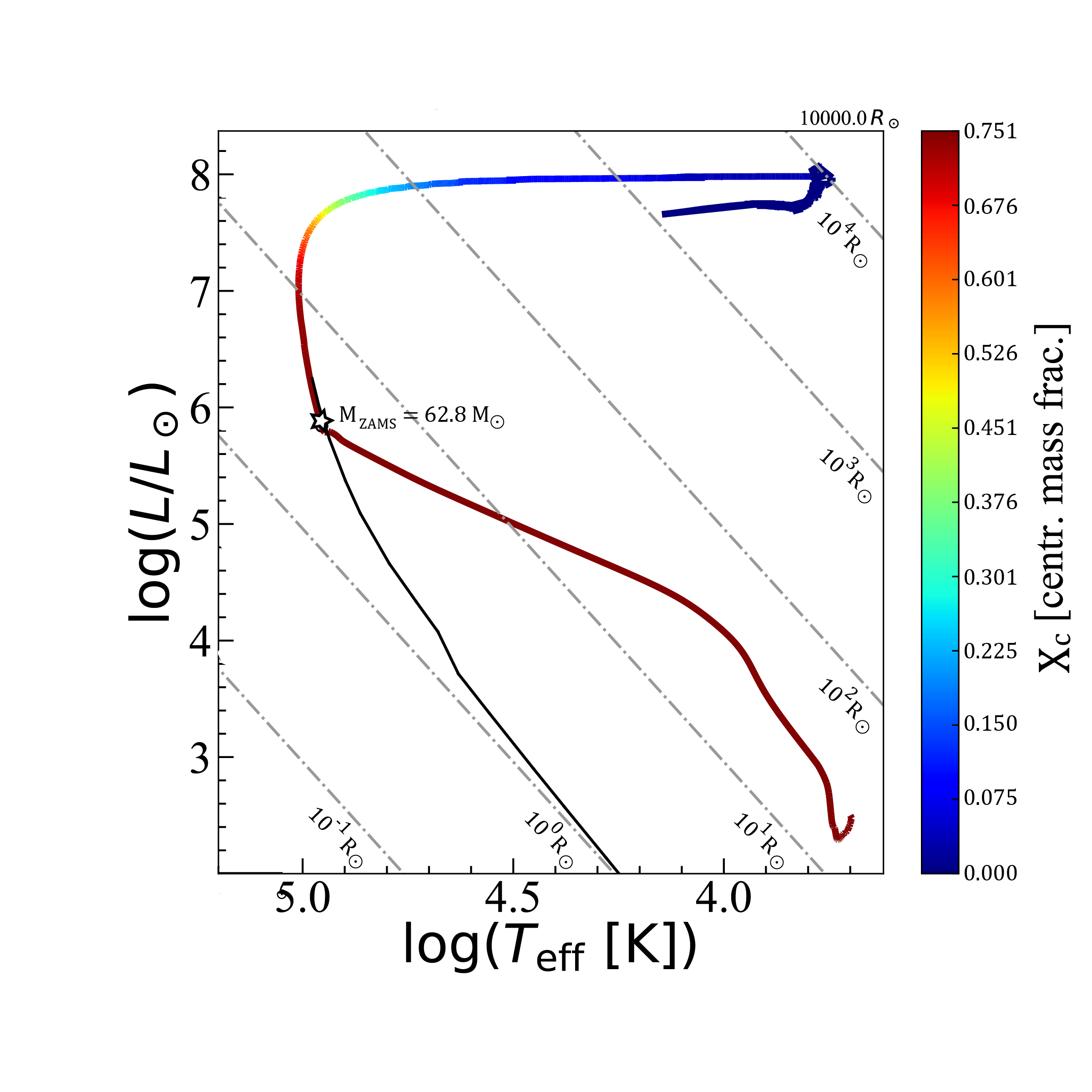}
                \includegraphics[width=9cm]{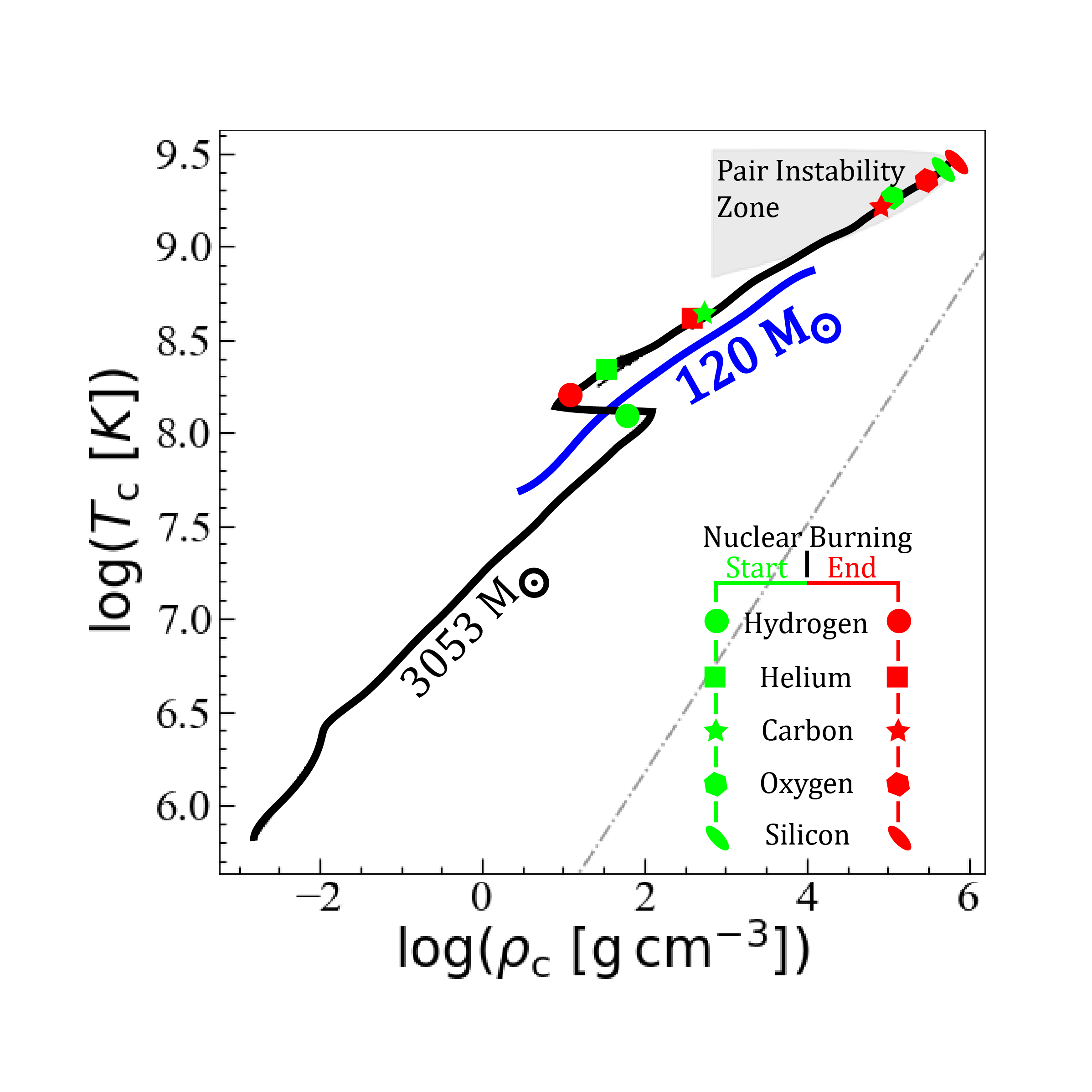}
        \caption{Evolution of a 3000 M$_\odot$ model at a constant accretion rate. Left panel: HR diagram depicting the evolution of a model until the end of core silicon burning with a constant accretion rate of $10^{-3}$M$_\odot$/yr. The black line represents the ZAMS line for Z=0 metallicity. The star symbol on the track represents the start of core hydrogen burning when the mass of the model is 62.8 \msun. Right panel: Evolution of the central temperature versus the central density of the model. The blue line marks a 120 \msun model at Z=0 metallicity, starting from ZAMS and ending at core helium burning.}
                \label{Fig:HRD}
\end{figure*}

\begin{figure*}
        \centering
                \includegraphics[width=18cm]{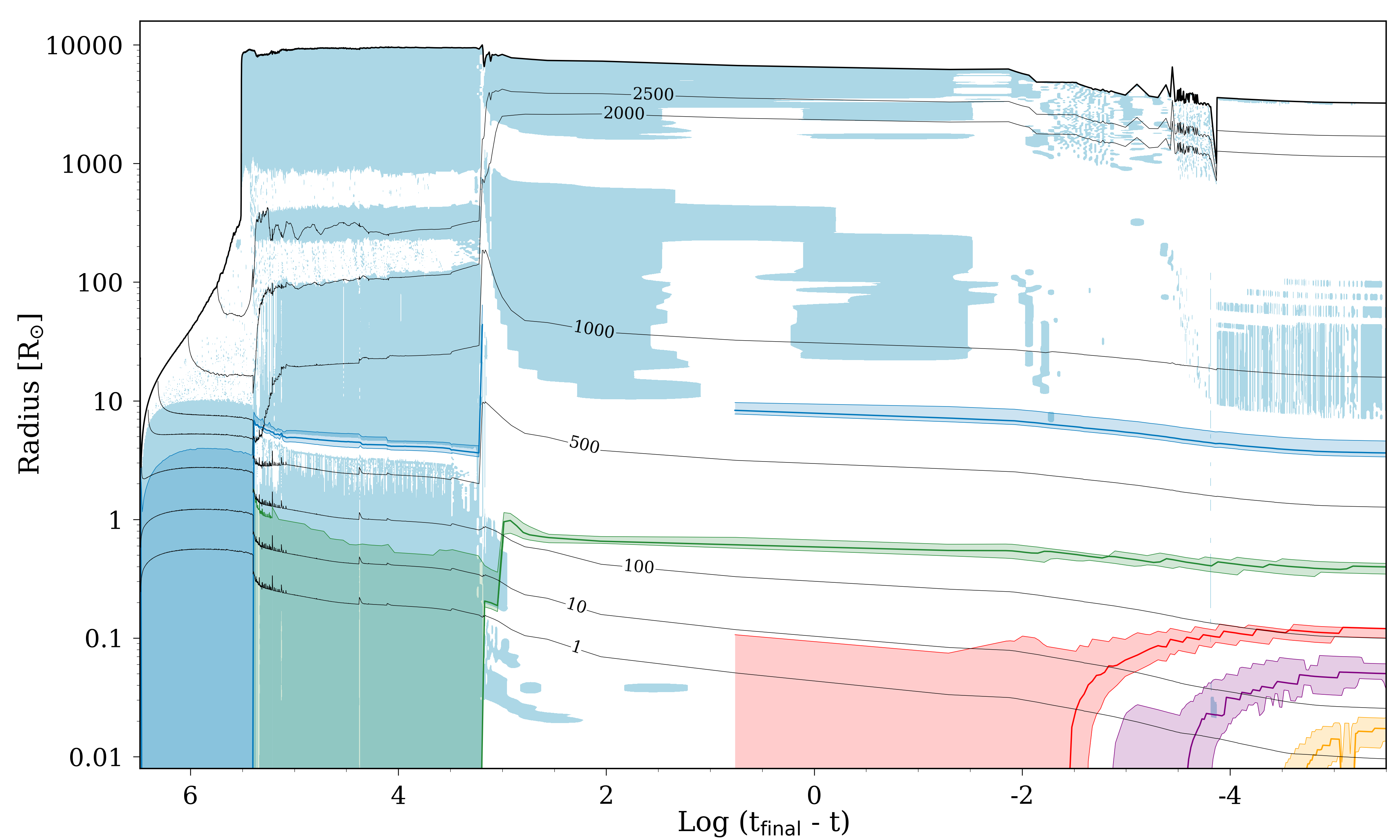}
        \caption{Internal evolution of the accreting model. The final mass is slightly above 3000 \msun. The convective regions are shaded in light blue. Shaded darker blue, green, red, purple, and orange regions indicate regions of hydrogen, helium, carbon, oxygen, and silicon burning, respectively. Iso-contours of mass are indicated by black lines.}
                \label{Fig:Kippen}
\end{figure*} 

\begin{figure}
        \centering
                \includegraphics[width=9cm]{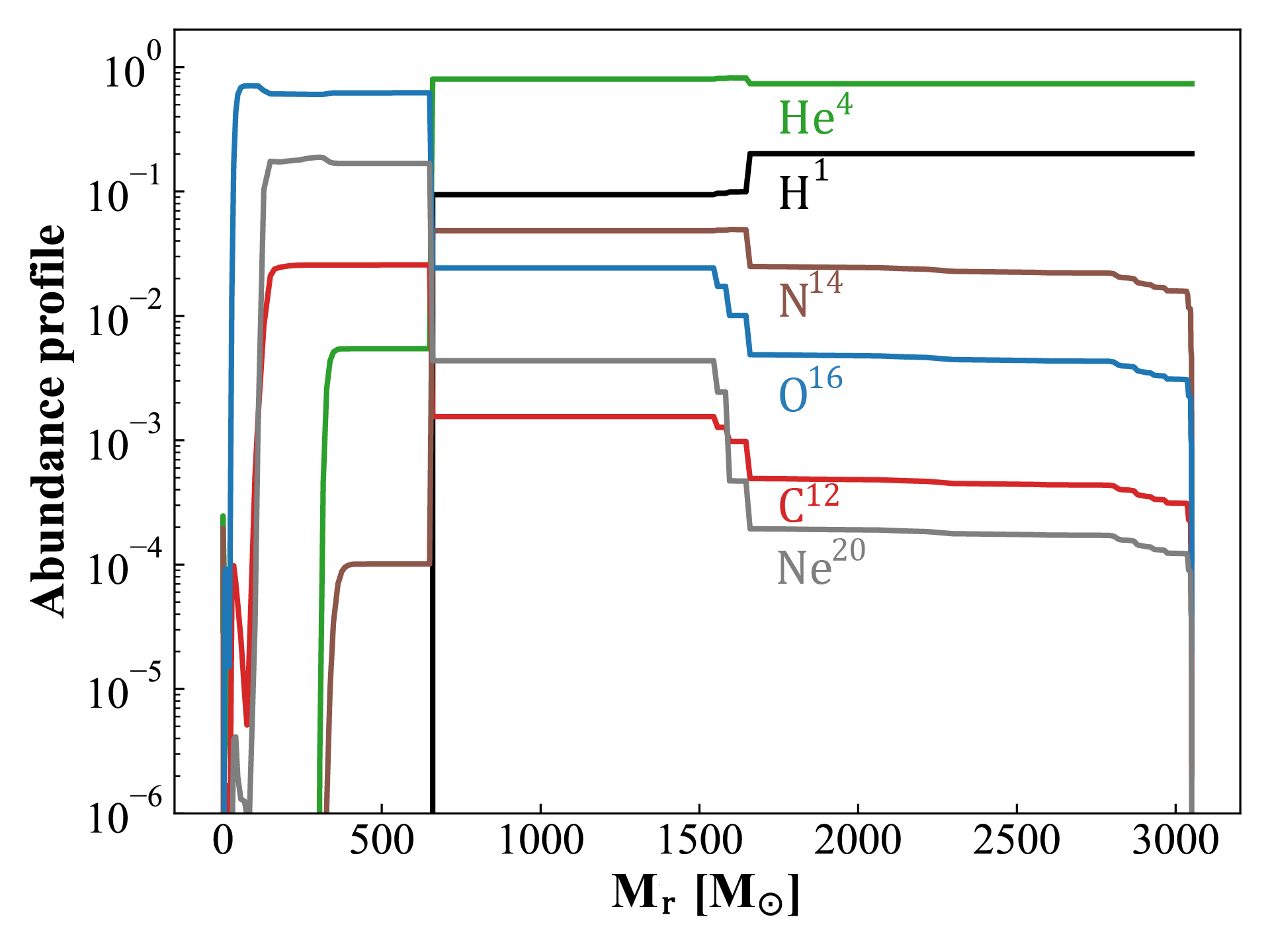}
        \caption{Abundance profile versus the mass coordinate of the zero-metallicity model with a constant accretion rate of $1\times10^{-3} M_{\odot}/yr$ at the end of core silicon burning.}
                \label{Fig:Abund}
\end{figure}

In this section we describe the evolution of a zero-metallicity star with a constant accretion rate of 10$^{-3}$ M$_\odot$ per year.
The evolution of our model in the Hertzsprung-Russell (HR) diagram and the variation in the central temperature and density are shown in Fig. \ref{Fig:HRD}.
Table \ref{tabModels} presents various physical quantities at the end of each nuclear burning phase.
We define the beginning of each nuclear burning phase as the point at which $10^{-3}$ of the mass fraction in the core of the main fuel at the beginning of each nuclear burning phase has been burned, and the end when $< 10^{-3}$ is left in the core.

\subsection{Pre-main-sequence phase}

The model begins as a 2 M$_\odot$, fully convective (central temperature T$_{c}=6\times10^{5}K$), and  chemically homogeneous seed (initial abundances in mass fraction of H$^{1}=0.7516$, He$^{4}=0.2484$, and H$^{2}=5.10^{-5}$) and subsequently grows in mass via accretion.
Once the central temperature exceeds 1.6$\times10^{6}K$, deuterium ignites in the core; however, the energy produced from deuterium burning has no impact on the evolution. 
This phase ends when hydrogen begins to be transformed into helium at the centre. At this stage, the central temperature is 1.26$\times10^{8}$K and the mass is 62.8 M$_\odot$.

\subsection{Core hydrogen burning}

The evolution during the core hydrogen burning can be divided into two phases. The first phase corresponds to a growth of the convective core in both mass and radius\footnote{The initial mass of the convective core is higher at a given stage during the core-hydrogen-burning phase. Here the total mass is increasing and thus, as expected, the convective core is increasing as well.}. The model increases in luminosity from log $L/L_{\odot}$=5.85 to 7.45 while maintaining a near-constant effective temperature. The stellar radius follows a monotonic relationship, with mass given by the power law $R \propto M^{1/2}$, as described by \cite{Hosokawa_2012}.
The second phase is marked by a reduction in the growth rate of the convective core.
Due to the decrease in hydrogen in the convective burning core, the nuclear timescale becomes shorter, and thus smaller quantities of mass have time to be accreted. 
When the convective core stops growing in radius, the star evolves towards lower effective temperatures at nearly constant luminosity. 
When core hydrogen is depleted, the star has a mass of 2803 M$_\odot$ and an effective temperature of $\mathrm{\log T_{eff}}= 3.75$.
The central temperature is 1.77$\times10^{8}$K, 40\% higher than the central temperature at the start of core hydrogen burning.

The evolution of such an extremely massive model is quite different when compared to a more classical massive star model in the log T$_{c}$ versus log $\rho_{c}$ diagram (see the right panel of Fig.~\ref{Fig:HRD}). In the classical non-accreting model, both the temperature and density increase from ZAMS to the end of core helium burning.  In the case of an extremely massive model, however, the temperature increases very slightly and the density decreases during the core-H-burning phase. This effect can be attributed to the substantial increase in mass due to accretion (62.8 M$_\odot$ at the start of H burning to 2$\,$800M$_\odot$ at the end of core H burning), which also leads to the growth in mass and radius of the convective core. We note that the duration of the core-H-burning phase is very similar to that of a 120 M$_\odot$ Pop III star\footnote{When the mass of a star increases, a larger fraction of the total pressure is due to radiation pressure. This impacts the mass-luminosity relation, which becomes flatter as the mass increases. Since the lifetime scales as $M/L$,, this means that above some mass, the lifetime no longer changes much when the mass increases.}

\subsection{Core helium burning}

After the end of core hydrogen burning, contraction of the core leads to a rise in central temperature, and helium is ignited in the core at 2.80x$10^{8}K$.
During core helium burning, several intermediate convective zones are formed.
Along with the apparition of an outer convective zone, this transitions the model into a near-fully convective structure, which facilitates the transport of helium from the core to the surface  (see Table \ref{tabModels}). Additionally, carbon is transported from the core to the hydrogen burning shell, where it is processed in the CNO cycle. This produces primary nitrogen, which is transported to the surface via these convective zones. 
This opens the possibility of production of primary nitrogen, which may significantly impact chemical feedback.

The model follows a more classical evolution along a 1/3 slope in the log T$_{c}$ log $\rho_{c}$ diagram (Fig. \ref{Fig:HRD}) because the growth in mass compared to the total mass is more moderate than during core hydrogen burning.
In addition, the monotonic relation between the stellar radius and mass is replaced by a nearly flat evolution of the radius. 

During this phase, the star is a red supergiant with a radius of 10$\,$000 R$_\odot$, of the order of ten times larger than the most luminous red supergiants \citep{Meynet2015}.
The duration of the core helium burning phase is 9\% of the core hydrogen burning phase, similar to a 120 M$_\odot$ model.

\subsection{Late evolutionary stages}

At the end of core helium burning, the central carbon abundance is 0.025. Before central carbon burning begins, the central conditions are in the pair-production zone ($\Gamma_1<4/3$), with 10\% of the CO core and 2\% of the total mass inside. This phase lasts for 12 years, which is similar to the core carbon burning of a 120 M$_\odot$ star.
The final burning stages take place in a radiative core.
Photo-disintegration of neon is the next evolutionary phase and occurs over a span of a few days. 
Core oxygen burning is next and lasts for less than a day; during this stage, 56\% of the CO core mass (7\% of the total mass) is in the pair-production zone. Before the end of oxygen burning, the centre exits the pair-production zone. Core silicon is exhausted within half a year, and by the end of this final stage, 50\% of the CO core mass (9\% of the total mass) is in the zone. The star may directly collapse with no ejection and produce a black hole of 3000 M$_\odot$.
Alternatively, part of the envelope may be ejected, leaving behind a black hole with a mass between 660 and 3000 \msun. However, the physics of such an explosion is complex, and detailed hydrodynamical models would be needed to assess the detectability of these supernovae and their chemical, mechanical, and radiative feedback \citep[e.g.][]{Ohkubo2009}.
The total luminosity output of models during the late evolutionary stages is dominated by the energy transported by neutrinos. The neutrino luminosity is more than a million times the photon luminosity at the end of silicon burning.
Applying the criteria from \citet{Lionel2021}, we find that the model never reaches the general relativistic instability.

We find that the most massive star that can be formed with a constant accretion rate of 10$^{-3}$ M$_\odot$ per year is slightly more than 3000 M$_\odot$. Unless some process as an additional source of energy were accounted for \citep[e.g. the disintegration of weakly interactive massive particles studied in][]{Freese2008,Taoso2008}, this would represent a robust upper mass limit. Most of the mass is obviously accumulated during the longest hydrogen nuclear burning phase.

\begin{table}[h]
\caption{Selected quantities from the evolutionary model at the end of each burning stage.}
\label{tabModels}
  \centering
\resizebox{\linewidth}{!}{
\begin{tabular}{l | c | c | c | c | c | c }
\hline
\textbf{Stage} & Mass & Duration & Log T$_{\mathrm{eff}}$ & log(g) & Y$_{\mathrm{surf}}$ &  M$_{\rm CO}$ \\

 & M$_\odot$ &  yrs  & & & & M$_{\odot}$ \\ 
\hline \hline

$^{2}$H      &$5.7$&5.8$\times10^{4}$& $3.74$& $2.81$& $0.25$& $0.00$ \\
$^{1}$H     &$2801$&2.7$\times10^{6}$& $3.76$& $-0.09$& $0.29$& $0.00$\\ 
$^{4}$He     &$3052$&2.4$\times10^{5}$& $3.77$&$-0.09$&$0.58$& $682$ \\
$^{12}$C     &$3052$&1.2$\times10^{1}$& $3.77$&$-0.08$&$0.59$& $673$ \\
$^{20}$Ne    &$3053$&9.9$\times10^{-3}$& $3.79$&$0.18$&$0.74$& $660$ \\
$^{16}$O    &$3053$&1.9$\times10^{-3}$& $3.94$&$0.88$&$0.74$& $660$ \\
$^{28}$Si    &$3053$&5.8$\times10^{-5}$& $3.94$&$0.88$&$0.74$& $660$ \\

\hline
\end{tabular}
}
\end{table}

\section{Impact of massive Pop III stars}\label{Sec:Impact}

To estimate the impact of massive Pop III stars, we needed to estimate their numbers with respect to classical massive stars
Assuming an initial mass function given by \citet{Kroupa2001} and \citet{Chabrier2003}\footnote{The number of stars born with an initial mass M is proportional to M$^{-(1+x)}$, with $x=1.3$ for stars with initial masses larger than 1 M$_\odot$.}, one 3000 M$_\odot$ star is expected to form for every 10$^5$ 20 M$_\odot$ stars. 
This value depends on the choice of initial mass function. For instance, choosing a top-heavy initial mass function as used in, for example, \citet{Baugh2005} would change that to 150, significantly enhancing the impact of such extremely massive stars.

\subsection{Chemical feedback}\label{Sec:Chemistry}

Figure \ref{Fig:Abund} shows the chemical structure at the end of core silicon burning. During core helium burning, many intermediate convective zones are formed in the star, and these zones transport helium from deep in the stellar interior to the surface. This results in significant surface helium enrichment, with a final value of 0.74 compared to the initial value of 0.25.
This value is atypical for a non-rotating star without mass loss.
In addition, the regions above the CO core are also strongly enriched in nitrogen, oxygen, and, to a lesser extent, carbon and neon. Through their winds or via mass ejection at the time of the core collapse, these stars may be significant enrichment sources for their surroundings.

Considering the extreme case where the whole envelope above the CO core is lost, the ejected material would consist of 2400 M$_\odot$ of helium, 72 M$_\odot$ of $^{14}$N, 24 M$_\odot$ of $^{16}$O, 2.4 M$_\odot$ of $^{12}$C, and 2.4 M$_\odot$ of $^{20}$Ne.
In comparison, the 20 M$_\odot$ Pop III evolutionary model from \citet{Murphy2021} would eject 1.4 10$^{-3}$ M$_\odot$ of $^{14}$N, 0.64 of $^{12}$C, and 1.64 M$_\odot$ of $^{16}$O.
Therefore, one 3000 M$_\odot$ star would eject a quantity of $^{14}$N equivalent to more than 50 000 20 M$_\odot$ Pop III stars, contributing as much as half the quantity produced by a whole population of 20 M$_\odot$ stars. However, the contribution is less significant for other elements.

\citet{Piau2006} suggested that a first generation of massive stars may have caused the lithium to be depleted in the interstellar medium and have evoked that possibility to explain the difference between the cosmological lithium abundance and the one deduced from the `Spite' plateau lithium \citep{Peri2022}.
We also note that lithium is depleted in the envelope of our 3000 M$_\odot$ star. However one such star would participate in the lithium destruction in an ejected mass that might be at most one-thousandth of the mass ejected by a population of 20 M$_\odot$ stars.
A point, however, that may still make such stars interesting objects in that context is the fact that if only the envelope were ejected, the mass would be iron-free. Thus, such stars would contribute to depleting lithium from the interstellar medium while not enriching it in iron. This question needs to be studied in greater detail.

Stars that formed from the envelope material of a 3000 M$_\odot$  red supergiant would be helium-rich, lithium-depleted, and exhibit high N/C and N/O ratios.
They would thus present a composition characteristic of both CNO and triple alpha enrichment.
The N/C and N/O ratios obtained in the envelope of our model at the end of its evolution are 30 and 3, respectively.
This may be similar to the CNO abundances observed at the surface of very iron-poor stars. For example, the carbon-enhanced metal-poor star HE1327-2326 \citep{Collet2006}, with a [Fe/H] of~-5.6, shows nitrogen abundances larger than those of carbon and oxygen, although with values of 2.5 for N/C and 6 for N/O. Further study, including of the explosive nucleosynthesis, is needed to investigate whether extremely massive stars contribute to such chemical signatures.

The recent detection of bright NIII and NIV emission lines in GN-z11 points towards a high nitrogen enrichment in this galaxy \citep{Bunker2023}. The value of N/O (log(N/O) $> -0.25$) for GN-z11 is found to be four times higher than solar values \citep{Cameron2023,Senchyna2023}. Works by \citet{Charbonnel2023} and \citet{Nagele2023} have attempted to explain SMSs as the main polluter responsible for the high nitrogen enrichment. Our zero-metallicity model has values of log(N/O)$= 0.45$ and log(C/O)$= -1.01$, assuming the entire envelope above the CO core is ejected. Depending on the fraction of the envelope that is ejected, these values could vary between $0.45 < $ log(N/O) $< 0.75$ and $-1.01 < $ log(C/O) $< -0.87$.
For the model with Z = $10^{-6}$, we obtain log(N/O)$\approx 0.55$ and log(C/O)$\approx -0.93$.
The value of N/O is above the lower limit for GN-z11 reported by \citet{Cameron2023}, while the value of C/O is slightly below the lower limit.
This suggests that accreting Pop III stars of $\sim$ 3000 \msun could be the cause of the high N/O abundance observed in GN-z11.
However, future work is required to perform a more detailed comparison with GN-z11 and any future objects that are detected by JWST.

\begin{figure}
        \centering
                \includegraphics[width=8cm]{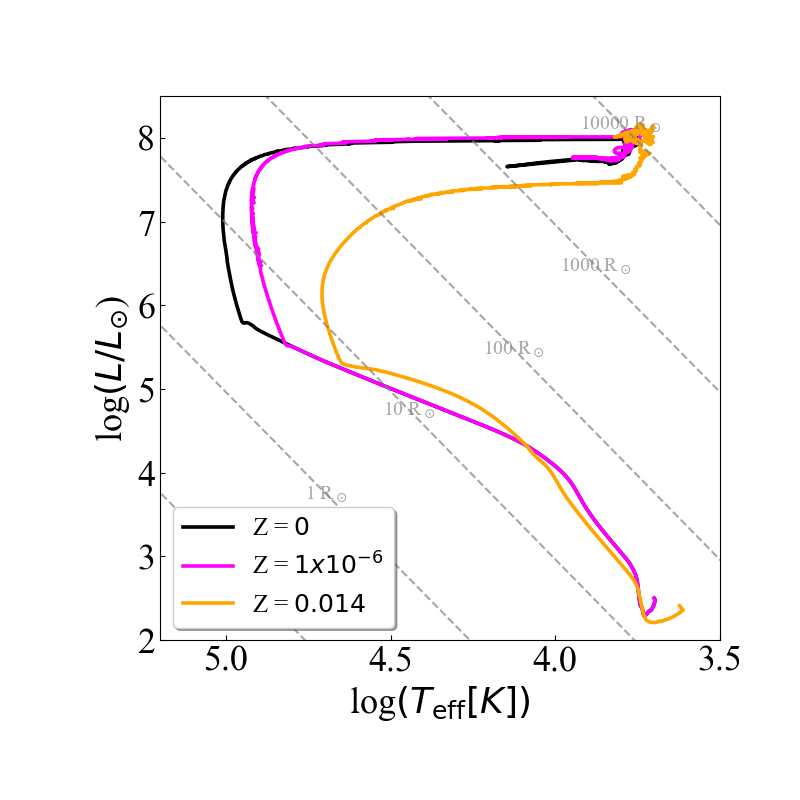}
        \caption{Evolution of three models computed using a constant accretion rate of $10^{-3}$ M$_\odot$/yr until the end of core silicon burning. The black line represents Z=0, the magenta line Z=$1\times10^{-6}$, and the orange line Z = 0.014. The iso-radius lines are represented in grey.}
                \label{Fig:Metalcomp}
\end{figure}

\subsection{Mechanical and radiative feedback}

Mechanical energy can be ejected by such a 3000 M$_\odot$ through stellar winds and/or at the time of the supernova explosion.
The escape velocity of the star in the red supergiant phase at the end of core helium burning is $\sim$ 340 km/s. This is about four times larger than the escape velocity of a 20 M$_\odot$ red supergiant (around 90 km/s). 
Assuming 1000 M$_\odot$ were lost via a stellar wind in the red supergiant phase (requiring a mass-loss rate of several times 10$^{-3}$ M$_\odot$/yr), the total mechanical energy ejected by winds would be about 1400 times the mechanical energy of a 20 M$_\odot$ red supergiant losing about 10M$_\odot$.
Under these assumptions, one 3000 M$_\odot$ star would inject a mechanical energy of about 1\% of the whole population of 20 M$_\odot$ stars.
One massive 3000 M$_\odot$  star has a luminosity that is 100-1000 times that of one Pop III 20 M$_\odot$ star. However, the contribution to the radiative feedback from a single 3000 \msun star dwarfs in comparison to a population of 10$^5$ 20 M$_\odot$ stars.
Extremely massive stars would also not be the key agents for producing ionising photons. 
This comes from the fact that in high-mass stars, the number of ionising photons per unit mass remains approximately constant \citep[see e.g.][]{Sibony2022}. As long as the same amount of mass is locked in massive stars, the ionising photon budget is not significantly affected by whether the mass is locked in more numerous, less massive stars or in less numerous, more massive stars. Also, intermediate mass black holes of $\sim 100 - 1000$ M$_\odot$ can be formed from extremely massive stars and be sources of high-energy photons if they continue to accrete material.

\subsection{Comparison with previous work}\label{Sec:Adv}

\begin{figure*}[!t]
        \centering
                \includegraphics[width=9.3cm]{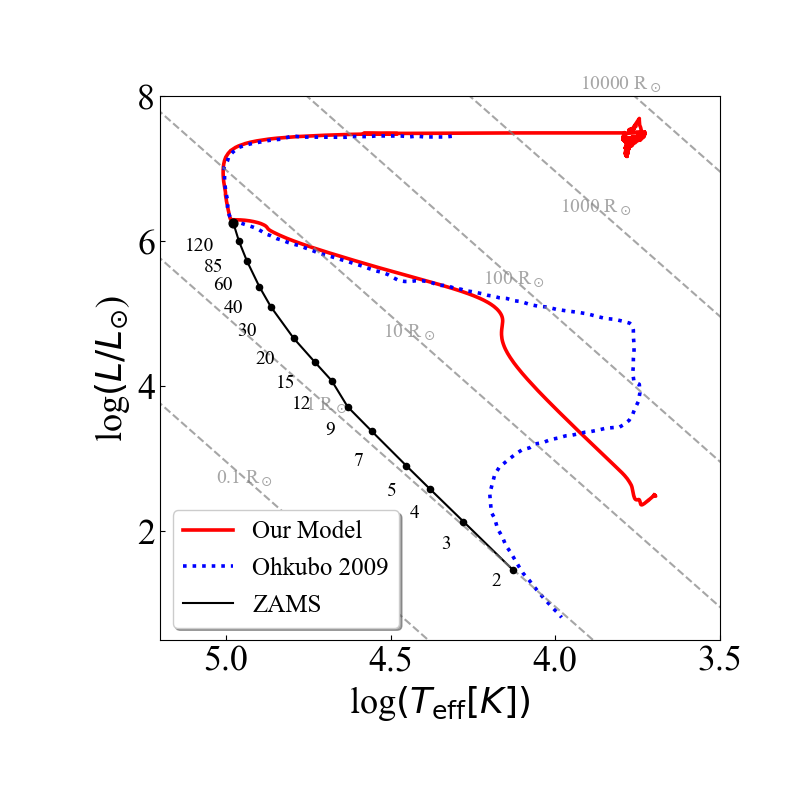}\includegraphics[width=9.3cm]{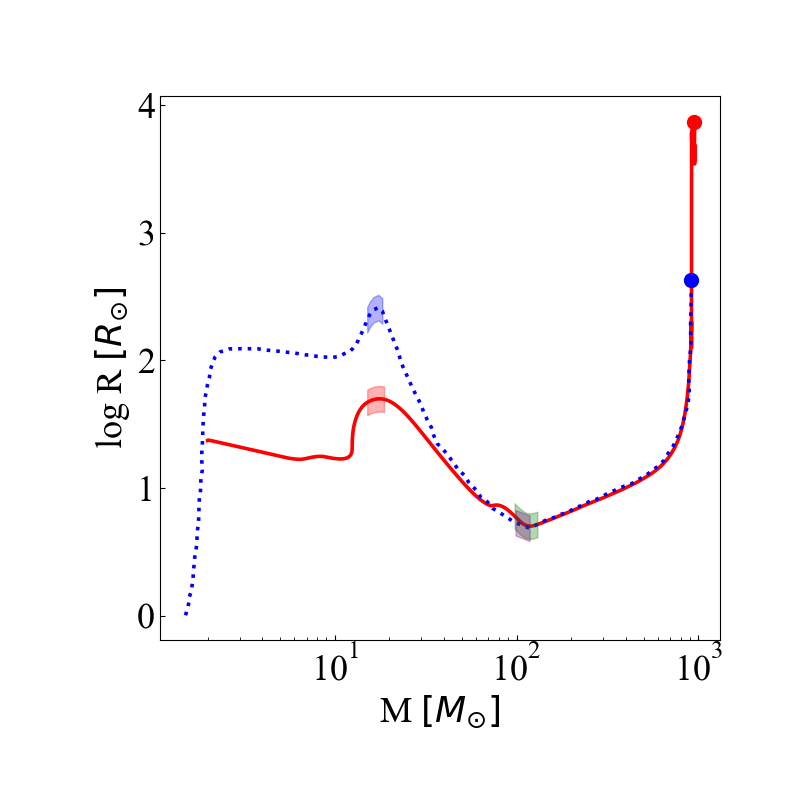}   
        \caption{Evolution of accreting Pop III models following the accretion law of \citet{Yoshida2007} depicted in the HR diagram. The dotted blue line corresponds to the Y-1 model in Fig. 5 of \citet{Ohkubo2009}, and the solid red line represents the model computed in this work using \gva. The black line represents the ZAMS line, and the dots indicate the masses in M$_\odot$. The dotted grey lines are the iso-radii, with values depicted in solar radii.{\it Right panel:} Evolution of the radius versus mass of models shown in the left panel. Due to different initial conditions, the two models possess different radii during the pre-main sequence (first shaded region) and reach core hydrogen burning at a near-identical mass of around 120 M$_\odot$ (second shaded region). A stark difference arises at the core Si burning, as modelled by \citet{Ohkubo2009}, which finishes the evolution in blue at a radius of log(R) $\approx$ 2.5 (blue dot), whereas our model reaches the Hayashi limit with log(R) $\approx$ 4 (red dot).}
                \label{Fig:Yosh}
\end{figure*}

\citet{Ohkubo2009} computed evolutionary models of accreting massive stars. 
They adopted a mass-dependent accretion rate from the cosmological simulations of \citet{Yoshida2007}.
In Fig.~\ref{Fig:Yosh} we compare our \citet{Yoshida2007} model with theirs.
Our model starts with a higher initial mass of 2 M$_\odot$ with $\log (L/L_{\odot})=2.46$ and $\log (T_{\rm eff}) = 3.69$, whereas the model in \citet{Ohkubo2009} starts at 1.5 M $_\odot$ with $\log (L/L_{\odot})=0.80$ and $\log (T_{\rm eff}) = 3.98$. 
Additionally, the evolution of the radius as depicted in the right panel of Fig.~\ref{Fig:Yosh} shows that our proto-stellar seed has a much larger radius, $19~R_\odot$  compared to the $\approx 1~R_\odot$ of \citet{Ohkubo2009}.
However, this has little impact on the subsequent evolution.

After a pre-main-sequence evolution of ~$\approx 10^{5}$ years (see the second shaded region in the right panel of Fig.~\ref{Fig:Yosh}), both models finish the pre-main-sequence evolution at 120 M$_\odot$ and begin core hydrogen burning. 
The evolution of luminosity during this phase is very similar, but as the models approach the end of core hydrogen burning, our model has a lower effective temperature, $\log (T_{\rm eff}) = 4.47$  compared to $\log (T_{\rm eff}) = 4.90$ for \citet{Ohkubo2009}. 
This is possibly due to the different treatment of energy transport in the outer layers of the two models.
The evolution beyond core hydrogen burning shows a stark difference: our model migrates to the red with an effective temperature of $\log (T_{\rm eff}) = 3.73,$ whereas the model from \citet{Ohkubo2009} finishes core silicon burning at a much hotter $\log (T_{\rm eff}) = 4.32$. 
The two models may therefore exhibit a small difference in their ionising power and radiative and chemical feedback.

\section{Impact of varying metallicity}\label{Sec:Metal}

Supermassive stars have usually been thought to form in metal-free environments (Z = 0).
However, the formation of SMSs in metal-enriched, atomically cooled halos aided by super-competitive accretion has been explored by \citet{Chon2020}. 
Additionally, recent work by \citet{Hirano2022} suggests that the presence of magnetic fields during the accretion phase may allow the formation of SMSs at a metallicity of $Z/Z_{\odot} = 10^{-5}$.
To study the impact of metallicity on the evolution of $\sim 3000 M_{\odot}$ stars, we compared models with a constant accretion rate of $10^{-3}$ M$_\odot$/yr computed until the end of core silicon burning with metallicities of Z = 0, $10^{-6}$, and 0.014 (Fig.~\ref{Fig:Metalcomp}).
Each model has a similar pre-main-sequence evolution and reaches the ZAMS at 63 M$_\odot$, 58 M$_\odot$, and 61 M$_\odot$ for 0, $10^{-6}$, and 0.014, respectively.
Similarly to massive stars, a higher metallicity results in a larger radius and lower $T_{\rm eff}$  during the main sequence due to the higher abundance of CNO elements in the core and higher opacities.
A higher metallicity also results in a faster growth of the convective core mass during the main-sequence phase, resulting in a longer lifetime, a lower central temperature, and therefore a higher final mass. Models with Z~$ = 0$, $1\times10^{-6}$, and 0.014  have final masses of 3053, 3285, and 3478 M$_\odot$, respectively. 
When comparing the abundance profiles inside the three models at the end of the core-carbon-burning phase, we find that, due to the presence of a large number of intermediate convective zones in the outer 50\% (M$_r$/M$_{tot}$) of the 0 and $10^{-6}$ metallicity models, hydrogen and helium are fully mixed in these regions, totalling 294 M$_\odot$ (10\% of the total mass) and 1138 M$_\odot$ (37\% of the total mass), respectively. For the solar metallicity case, this region only extends to the outer 15\% of the model, where the mass fractions of hydrogen and helium are 166 M$_\odot$ (4.7\% of the total mass) and 331 M$_\odot$ (9.5\% of the total mass), respectively.
In summary, our models show that changing the metallicity has an impact on the final mass and on the mass and distribution of helium in the envelope.

\section{Conclusion}\label{Sec:Conc}

We have presented evolutionary models of accreting $\sim 3000 \msun$ stars to investigate their evolution and their chemical, radiative, and mechanical feedback on their surroundings.

\begin{itemize}

\item After core hydrogen burning, our models exhibit several intermediate convective zones below the convective envelope. This allows strong mixing of chemical elements inside the star and produces helium-rich red supergiants, with a final surface helium abundance of 0.74. The strongly mixed chemical structure makes these stars a potentially interesting source for nucleosynthesis. The star remains in the red region of the HR diagram during all subsequent burning phases.

\item By comparing the contribution of one 3000 M$_\odot$ star to a population of 20 M$_\odot$ stars, we find that one 3000 M$_\odot$ star may significantly contribute to primary nitrogen production. However, it does not have a significant impact on mechanical and radiative feedback compared to the population of 20 M$_\odot$ stars.

\item The models do not reach the general relativistic instability. However,  central regions enter the electron-positron pair production region during oxygen burning. Hydrodynamical models are needed to compute what happens after the core collapse.

\item The maximum mass that can be reached by a constant 10$^{-3}$ M$_\odot$ star per year is around 3000 M$_\odot$. This limit is given by the nuclear lifetime.

\item Our models at Z = 0 and Z = $10^{-6}$ could pollute the interstellar medium with enriched abundances of $0.45 < $ log(N/O) $< 0.75$ and $-1.01 < $ log(C/O) $< -0.87$, depending on the fraction of the envelope that is ejected. This suggests that accreting 3000 \msun stars could help explain recent observations of the high-redshift galaxy GN-z11 \citep{Cameron2023}.

\end{itemize}

Future work could consider the impact of detailed accretion rate histories derived from cosmological simulations.
Any interruption of the accretion process may lead to significant mass loss, which would change the subsequent evolution of the star.
It would also be interesting to study the pulsational stability of such objects, and to investigate the impact of different treatments of convection, rotation, and magnetic fields. 

\section*{Acknowledgements}

D.N., E.F. and G.M. have received funding from the European Research Council (ERC) under the European Union's Horizon 2020 research and innovation programme (grant agreement No 833925, project STAREX). G.B. is funded by the SNF AMBIZIONE grant No 185805 (Seismic inversions and modelling of transport processes in stars). EF and GM  has received funding from SNF grant No 200020\_212124.

 \bibliographystyle{aa}
\bibliography{biblio}

\end{document}